\documentclass[aip]{revtex4-1}
\usepackage{graphics}
\usepackage{amsbsy}
\usepackage{amsmath}
\usepackage{amssymb}
\usepackage[dvips]{graphicx}
\usepackage{times}
\usepackage{balance} 
\usepackage{lastpage}
\usepackage[format=plain,justification=raggedright,singlelinecheck=false,font=small,labelfont=bf,labelsep=space]{caption} 
\usepackage{fancyhdr}

\draft % marks overfull lines with a black rule on the right

\begin{document}

%\preprint{}

\title{Dynamics of pearling instability in polymersomes: the role of shear membrane viscosity and spontaneous curvature} %Title of paper



\author{J. Lyu}
\affiliation{Aix Marseille Univ, Centrale Marseille, CNRS, M2P2, Marseille, France}`
\affiliation{Univ. Grenoble Alpes, CNRS, LRP, Grenoble, France}
%\thanks{}
\author{K. Xie}
\affiliation{Aix Marseille Univ, Centrale Marseille, CNRS, M2P2, Marseille, France}
\affiliation{Univ. Bordeaux, CNRS, LOMA, Bordeaux, France}
\author{R.  Chachanidze}
\affiliation{Univ. Grenoble Alpes, CNRS, LRP, Grenoble, France}
\author{A.  Kahli}
\affiliation{Aix Marseille Univ, Centrale Marseille, CNRS, IRPHE, Marseille, France}
\author{G.  Bo\"{e}dec}
\email[]{boedec@irphe.univ-mrs.fr}
\affiliation{Aix Marseille Univ, Centrale Marseille, CNRS, IRPHE, Marseille, France}
\author{M. Leonetti}
\email[]{fanleo42@gmail.com}
\affiliation{Aix Marseille Univ, Centrale Marseille, CNRS, IRPHE, Marseille, France}
\affiliation{Univ. Grenoble Alpes, CNRS, LRP, Grenoble, France}
\affiliation{Aix Marseille Univ, CNRS, CINaM, Marseille, France}


\date{\today}

\begin{abstract}
The stability of copolymer tethers is investigated theoretically. Self-assembly of diblock or triblock copolymers can lead to tubular polymersomes which are known experimentally to undergo shape instability under thermal, chemical and tension stresses. It leads to a periodic modulation of the radius which evolves to assembly-line pearls connected by tiny tethers. We study the contributions of shear surface viscosity and spontaneous curvature and their interplay to understand the pearling instability. The performed linear analysis of stability of this cylinder-to-pearls transition shows that such systems are unstable if the membrane tension is  larger than a finite critical value contrary to the Rayleigh-Plateau instability, an already known or if the spontaneous curvature is in a specific range which depends on membrane tension. For the case of spontaneous curvature-induced shape instability, two dynamical modes are identified. The first one is analog to the tension-induced instability with a marginal mode. Its wavenumber associated to the most unstable mode decreases continuously to zero as membrane viscosity increases. The unexpected second one has a finite range of unstable wavenumbers. The  wavenumber of the most unstable mode tends to a constant as membrane viscosity increases. In this mode, its growth rate becomes independent of the bulk viscosity in the limit of high membrane viscosity and behaves as a pure viscous surface. 
\end{abstract}


\maketitle %\maketitle must follow title, authors, abstract and \pacs

\section{Introduction}
Polymersomes are drops bounded by a copolymer membrane \cite{Disher1999Science}. They result from the self-assembly of diblock copolymers in a structure already encountered when phospholipids  self-organize in a lipid bilayer. Copolymers are versatile associating biocompatibility and biodegradability to a wide chemical diversity \cite{Liu2013SM,dionzou2016comparison}.  Cross-linking between diblock copolymers confers an increasing rigidity \cite{Robbins2011SM}. Under UV illumination, polymersomes with asymmetrical bilayers made of a copolymer with a rod-like conformation can exhibit bursting induced by curling \cite{Mabrouk2009PNAS}. The self-assembly of triblock copolymers in a single monolayer provides an one more tool to design containers reserved to drug delivery for example. Polymersomes are promising vehicles in biomedical applications \cite{Disher2006ARBE}. 

Polymersomes have analog properties to lipid vesicles. Basically, at the thermodynamical equilibrium, their shapes are governed by bending energy and the two constraints on the inner volume and on the surface area which are both constant because of the impermeability and the incompressibility of the membrane contrary to a droplet. Polymersomes like vesicles exhibit an extended zoology of shapes \cite{Seifert1997,Yanagisawa2008PRL,Disher1999Science,Li2013SMb}. Notably, polymersomes and vesicles can have a cylindrical shape with an aspect ratio up to 100. Their diameters vary from several micrometers \cite{Reinecke2003Langmuir,Sinha2013SM} to several tenths of nanometers \cite{Hamley2005SM,Wang2014MacroLetters}. Speaking more generally on lipids and copolymers, these tubes can result from various physical origins: a spontaneous curvature, the fabrication leading to a cylinder-like steady state which corresponds to a local minimum of energy \cite{Wang2014MacroLetters}, the pulling by a local force from a mother vesicle \cite{Reiner2006PNAS,Smith2004PRL,Karlsson2001Langmuir,Powers2002PRE,Derenyi2002PRL} or the emergence of tubes from vesicles due to an external flow \cite{Dittrich2006LabChip,Kanstler2008PRL}, an electric field \cite{Sinha2013SM} or an osmotic   for examples. Sedimentation of vesicles \cite{Boedec2011JCP,Boedec2012JFM,Rey2013Plos} is a typical example of flow-induced deformations, from quasi-sperical shapes to very thin tubes varying the Bond number \cite{Huang2011NJP,Boedec2013PRE}. The process is analog in withdrawal configuration \cite{Chatkaew2009PRL}. Tubulation can also result from encapsulating biopolymers and osmotic deflation \cite{Okano2018} and from a mixing of membrane lipids \cite{Bhatia2018} for example.

The polymersome tubes can evolve to a modulated (or corrugated) shape under stretching \cite{Reiner2006PNAS} or after a thermal quench \cite{Reinecke2003Langmuir} resulting in a pattern reminiscent from the viscous Rayleigh-Plateau instability \cite{Tomotika1935} and the pearling instability along lipid tubes induced by an optical tweezer \cite{BarZiv1994PRL}, an electric field \cite{Sinha2013SM}, a magnetic field \cite{Menager2002EPJE}, the gravity field \cite{Huang2011NJP} or by interactions with anchored amphiphilic polymers \cite{Tsafrir2001PRL} or nanoparticles \cite{Yu2009JACS}. Note that pearling is also observed in biological cells under active processes \cite{Jelercic2015} but also when cortical actin network is lacking \cite{Heinrich2014BJ}. It is possible to distinguish the pearling instability in two sub-classes: tension-induced pearling and spontaneous curvature-induced pearling. Anyway, the pattern along polymersomes emerges on a very slow characteristic time compared to experiments on lipid tethers. The viscous dissipation in the membrane is expected to play a role. 
 
At a dynamical point of view, the membranes of polymersomes are characterized by their fluidity. More generally, fluid interfaces are characterized by several viscosities depending on their assembly structure: dilational and shear surface viscosities and intermonolayer friction. As self-organized copolymer membranes are incompressible, the dilational surface viscosity is not relevant in our case.  A membrane made of triblock copolymers is a monolayer and hasn't any contact friction contrary to the case of a bilayer of diblock copolymers, the most standard case. However, it is generally assumed that the intermonolayer friction plays a role in thin structures less than one micrometer in lipid membranes \cite{Seifert1993EPL,Honerkamp2013PRL}. Finally, for the stability study of copolymer tubes larger than several hundreds of nanometers, it is reasonable to consider an incompressible membrane with only the shear surface viscosity at least in a first approach.  

Contrary to lipid membranes, copolymer ones  present a very high shear resistance.  Their surface (or membrane) viscosity $\mu_s$ is approximately three orders of magnitude larger than lipid one \cite{Dimova2003EPJE,LeMeins2011EPJE}.  In their seminal work, considering the motion of a protein, P. G. Saffman and M. Delbr\"{u}ck introduced a characteristic length $L_{sd} = \mu_s/(\eta_i+\eta_o)$ where $\eta_{i,o}$ are respectively the bulk viscosities of inner and outer fluids \cite{Saffman1975PNAS, Saffman1976JFM}. The ratio $L_{sd}/R$ where R is the characteristic length of the system gives an order of magnitude of the dissipation inside the membrane versus the bulk one. It is called the shear Boussinesq number. If $L_{sd}/R << 1$, the effect of membrane shear is negligible. First, consider a fluid lipid bilayer embedded in water: $L_{sd} \approx 0.9$ $\mu$m in the disordered state and $L_o \approx 7.8$ $\mu$m in the ordered state deduced from the pattern of membrane flow of a vesicle adhered on a substrate  \cite{Honerkamp2013PRL}. For a mixing of DOPC, DPPC and cholesterol and varying the temperature, the range of $L_{sd}$ varies from $0.2$ $\mu$m to $100$ $\mu$m, values obtained by the measurement of the diffusion of lipid domains \cite{Cicuta2007JPCB,Petrov2008BiophysJ}. The membrane viscosity of Red Blood Cells is larger in the range $2-9$ $10^{-7}$ Pa.m.s which corresponds to a Saffman-Delbr\"{u}ck length about 30 $\mu$m \cite{Nash1983BiophysJ,Hochmuth1979BiophysJ,TranSonTay1984BiophysJ}. In outer air cells, $L_{sd}$ exceeds 1 mm \cite{Li2002BiophysJ}. For a polymersome made of PEO-PBd copolymer, $\mu_s \approx 4$ $10^{-6}$ Pa.s.m what means $L_{sd}\approx 1-2$ mm, a value determined by falling-ball viscosimetry and pulling a tether by optical tweezers \cite{Dimova2003EPJE,LeMeins2011EPJE}. All these experimental examples indicate that the contribution of shear membrane viscosity is expected to play a major role in the dynamics of polymersomes. A typical example is the transition between tank-treading and tumbling motion of a polymersome in a shear flow. The shear membrane viscosity should promote tumbling compared to tank-treading motion.  Surface viscosities play also a role in the shape and dynamics of surfactant-laden droplets \cite{LuoBai2019JFM,Gounley2016JFM}, capsules \cite{Bagchi2009PRE,Loubens2016JFM} and elastic tubes\cite{Bacher2021JFM,Mora2010PRL}.

In the present work, we study the stability of an initial cylindrical polymersome under an external forcing such as an applied surface mechanical tension or the appearance of a spontaneous curvature. The latter could be due to a change of copolymer conformation under illumination the former to an external force for example. A linear stability analysis is performed taking into account the shear membrane dissipation, a salient feature of polymersomes. In the following, the second part presents the modeling (bulk and boundary equations) used to describe the physical quantities such as pressure, flow velocity, tension and membrane velocity. Especially, the mechanical equilibrium at the membrane is detailed with each membrane force such as bending, tension and in particular the viscous force due to shear membrane viscosity. To be useful in any applications, the general expression of the viscous force is provided with the tools of differential geometry, the essential elements being recalled in appendix. The force is fully derived in the linear regime with the unexpected normal component. In the third part, the basic state is recalled. In the fourth part, all the equations are linearized and solved to study the behavior of perturbations of wavenumbers $k=2\pi/\lambda$ where $\lambda$ is the wavelength. In the fifth part, the dispersion relation $s\,=\,s(k)$ is determined, $1\,/\,|s|$ being the growing characteristic time of the perturbation $k$ if s is positive (unstable mode) or the damping time if s is negative (stable mode).

\section{Modeling}

\subsection{Problem}
\label{problem}
The model system is an infinite cylinder of an initial radius $R$ bounded by a membrane made of copolymers and embedded in an infinite bath. The inner and outer incompressible fluids are newtonian of viscosities $\eta_i$ and $\eta_o$ respectively. The aim of this study is to consider the stability of the membrane shape $r\,=\,R$ under a perturbation of the shape $\delta R(z,t)$ where $z$ is the coordinate along the cylinder axis and $t$ the time. This response is made quantitative by a linear analysis of stability of the system which provides  the dispersion relation $s\,=\,s(k)$ where $s$ is the growth rate and $k$ the wavenumber characterizing the space modulation of shape perturbation along the axis of the cylinder. If the perturbation $\delta R(z,t)$ increases ($s\,>\,0$) with time, the shape is unstable and stable in the contrary case ($s\,<\,0$).  

Various kinds of systems will be studied. In the case of two fluids on both sides of the membrane, the system is a polymersome with a membrane made of a monolayer of triblock copolymers or a bilayer of diblock copolymers. In the case of only one outer fluid, the system is a micelle with a membrane made of copolymers. The common characteristic of these systems is the high membrane viscosity compared to the lipidic one. The following results are established for the first case and extended to the second one.

In such a geometry, a point $\pmb{x}$ in the space is well defined by its cylindrical coordinates:
\begin{equation}
\pmb{x}\,=\,r(\theta, z)\,\pmb{e_r}\,+\,z\,\pmb{e_z}
\end{equation}
where ($\pmb{e_r};\, \pmb{e_\theta};\,\pmb{e_z}$) are the the unit vectors of the cylindrical basis. The aim of this study is to consider the stability of the membrane shape $r\,=\,R$ under an axisymmetrical perturbation of the shape $\delta R(z,t)$. A point $\pmb{x}_m$ at the interface is localized by:
\begin{equation}
\pmb{x_m}\,=\,(R\,+\,\delta R( z,t))\,\pmb{e_r}\,+\,z\,\pmb{e_z}
\label{xm}
\end{equation}

\begin{center}
\textbf{Bulk equations}
\end{center}

Considering typical parameter values, the density $\rho = 10^3$ kg.m$^{-3}$ and the viscosity $\eta = 1$ mPa.s of water, a tether radius $R \approx 1\, \mu$m  and a typical velocity $U \approx 1\, \mu$m.s$^{-1}$, inertial effects are negligible as the Reynolds number $R_e = UR\rho/\eta \approx 10^{-6}$. Thus, the pressure and the velocity field satisfy the Stokes equation:
\begin{equation}
-\pmb{\nabla} p^{i,o}+\eta_{i,o}\ \Delta \pmb{V}^{i,o} = \pmb{\mathit{0}}
\label{stokes}
\end{equation}
and the incompressibility equation:
\begin{equation}
\pmb{\nabla} .\ \pmb{V}^{i,o} = 0 
\label{divv}
\end{equation}
The pressure is an harmonic function. Indeed, the combination of the two previous equations (Eqs. \ref{stokes},\ref{divv}) leads to:
\begin{equation}
\Delta\, p^{i,o}\, =\, 0
\label{laplap}
\end{equation}

\begin{center}
\textbf{Boundary conditions}
\end{center}

The membrane made of copolymers is impermeable to the passage of solvent, small molecules and ions on the time scale of experiments. It involves that the normal component of bulk velocities is continuous: 
\begin{equation}
\pmb{V}^m\,=\,\frac{\partial \pmb{x_m}}{\partial t}
\end{equation}
\begin{equation}
\pmb{V}^m.\pmb{n}\,=\,\pmb{V}^{i}.\pmb{n}\,=\,\pmb{V}^{o}.\pmb{n}
\label{contVn}
\end{equation}
where $\pmb{x_m}$ is a point of the membrane and $\pmb{n}$ the outer vector normal to the membrane. If a membrane made of a single monolayer of copolymers is considered, the continuity of tangent velocities is necessarily satisfied. More generally, as explained in the introduction, a first reasonable approach of tubular polymersomes with a radius larger than several hundreds of nanometers is to consider the continuity of tangent velocity :
\begin{equation}
\pmb{V}^{m}.\pmb{t}\,=\,\pmb{V}^{i}.\pmb{t}\,=\,\pmb{V}^{o}.\pmb{t}
\label{contVt}
\end{equation}
where $\pmb{t}$ refers to the tangent vectors to the membrane.
If we consider a system made of a diblock copolymer bilayer, the characteristic time of flip-flop between the two monolayers is very large compared to the time of the experimental process. It is already the case for lipids which are much smaller molecules. Moreover, the compressibility coefficient of a copolymer monolayer is very high, larger than for lipids. The condition of surface incompressibility is satisfied:
\begin{equation}
\pmb{\nabla_s}.\pmb{V}^{i}\,=\,\pmb{\nabla_s}.\pmb{V}^{o}\,=\,0
\label{divvs}
\end{equation}
where $\pmb{\nabla_s}$ is the surface gradient.
Finally the membrane under flow is at the mechanical equilibrium:
\begin{equation}
(\pmb{\bar{\bar{\sigma}}}^o\,-\,\pmb{\bar{\bar{\sigma}}}^i)\,\pmb{n}\,+\,\pmb{f}^m\,=\,\pmb{0}
\label{equimec}
\end{equation}
where $\pmb{\bar{\bar{\sigma}}}\,=\,-p\pmb{1}\,+\,2\eta \pmb{\bar{\bar{D}}}$ is the newtonian stress tensor and $\pmb{f}^m$ is the mechanical membrane force per unit area. $\eta$ is the bulk viscosity and $\pmb{\bar{\bar{D}}}\,=\,(1/2)\,(\pmb{\nabla V}\,+\,\pmb{\nabla^T V})$ the strain rate tensor. Eq. \ref{equimec} provides the normal and tangent mechanical equilibriums.

\subsection{The membrane force}

\begin{center}
\textbf{The membrane viscous force}
\end{center}

The studied system concerns a surface that models a membrane of several nanometers of thickness. At a general point of view, a local parametrization $(s^1;\,s^2)$ is necessary to describe the variations of geometrical and physical quantities along the surface. The contravariant local basis $(\pmb{t}_1;\,\pmb{t}_2;\,\pmb{n})$ permit to separate the membrane velocity in its tangent and normal parts:
\begin{equation}
\pmb{V}^m\,=\,V^\beta\,\pmb{t}_\beta\,+\,V_n\,\pmb{n}
\label{Vbeta}
\end{equation}
where the notation of Einstein is used and $V^\beta$ is the contravariant coordinate. The general viscous interfacial force can be derived from the Scriven stress tensor \cite{Scriven1960}. Here, we consider a viscous membrane with a shear resistance and without any dilatational viscosity as the membrane is a bidimensional incompressible fluid. The expression of the force per unit area $\pmb{f}_v$ is more intricate than bending and tension forces \cite{Powers2010RMP}:
\begin{eqnarray}
\pmb{f}_v\,=\,\mu_s\,(\nabla^2V^\beta+KV^\beta-2V_n\nabla^\beta H)\:\pmb{t}_\beta\nonumber\\+2(Hg^{\alpha\beta}-K^{\alpha\beta})\nabla_\alpha V_n\:\pmb{t}_\beta \nonumber\\ \:+\:2\mu_s\Big(K_{\alpha\beta}\nabla^\alpha V^\beta\,-\,V_n(4H^2-K)\Big)\pmb{n}
\label{fv}
\end{eqnarray}
where $H$ is the mean curvature, $K$ the gaussian curvature, $g^{\alpha\beta}$ the metric tensor in the covariant basis and $K_{\alpha\beta}$ the contravariant curvature tensor. All these quantities are defined in the appendix with some additional elements of differential geometry to explain in details how to the derive some expressions obtained in the following parts.

\begin{center}
\textbf{The tension and bending force}
\end{center}

The response of a copolymer membrane to a constraint is governed first by the bending energy well described by the Helfrich energy, used extensively in the literature :
\begin{equation}
F_\kappa = \frac{\kappa}{2}\int_{S_m} \big[2H(\pmb{x}_m)-H_0\big]^2 dS
\end{equation}
where $\kappa$ is the bending modulus of the order of twenty thermal energy $k_BT$ of the membrane, $H_0$ its spontaneous curvature characteristic of the system and H the mean curvature. 
\begin{equation}
F_\gamma\, = \,\int_{S_m} \gamma(\pmb{x}_m) dS
\end{equation}
where $\gamma$ is the membrane tension associated with the $2D$ incompressibility of membrane flow. $\gamma$ is an unknown quantity equivalent to a bidimensionnal pressure  associated to the constraint of incompressibility (eq. \ref{divvs}). Thus there is no reason to have a constant tension in an unstationnary physical configuration. The forces per unit area are:
\begin{eqnarray}
\pmb{f}_\kappa\, =\,-\frac{\delta F_\kappa}{\delta \pmb{x}_m}\,=\, -\kappa(2\Delta_sH\,+\,4H(H^2-K))\pmb{n}\nonumber\\
-2\kappa H_0 K\pmb{n}\,+\, \kappa H_0^2 H \pmb{n}
\label{fkap}
\end{eqnarray}
\begin{equation}
\pmb{f}_\gamma\, =\,-\frac{\delta F_\gamma}{\delta \pmb{x}_m}\,=\, \pmb{\nabla}_s \gamma+\, 2\gamma H \pmb{n}
\label{fgam}
\end{equation}
The minus sign in eq. \ref{fkap} comes from our convention: the radius of curvature of a sphere is negative. Note that the last term of bending force can be recast in the tension force setting $\gamma\,=\,\bar\gamma-(1/2)\kappa H_0^2$. The total membrane force $\pmb{f}^m$ appearing in the mechanical equilibrium (eq. \ref{equimec}) satisfies:
\begin{equation}
\pmb{f}^m\,=\,\pmb{f}_\kappa\,+\,\pmb{f}_\gamma\,+\,\pmb{f}_v
\end{equation}

%%%%%%%%%%%%%%%%%%%%%%%%%%%%%%%%%%%%%%%%%%%%%%%%%%%%%%
%%%%%%%%%%%%%%%%%%%%%%%%%%%%%%%%%%%%%%%%%%%%%%%%%%%%%%
%%%%%%%%%%%%%%%%%%%%%%%%%%%%%%%%%%%%%%%%%%%%%%%%%%%%%%
%%%%%%%%%%%%%%%%%%%%%%%%%%%%%%%%%%%%%%%%%%%%%%%%%%%%%%

\section{The basic state}

In the basic state, the shape is a cylinder of radius R under tension $\gamma_0$. All the quantities are named by a superscript $(0)$ which means order zero to contrast with perturbations which are of the order one. There is no fluid flow:
\begin{equation}
\pmb{V}^{(i,o),(0)}\,=\,\pmb{0}
\label{v00}
\end{equation}
The external pressure $p^{o,(0)}$ is set to zero while the internal pressure $p^{i,(0)}$ is given by the normal component of the mechanical equilibrium with the effective membrane tension $\bar\gamma_0$:
\begin{equation}
p^{i,(0)}\,=\,p_0\,=\,\frac{\gamma_0}{R}-\frac{\kappa}{2R^3}\,+\,\frac{\kappa H_0^2}{R}\,=\,\frac{\bar\gamma_0}{R}-\frac{\kappa}{2R^3}
\end{equation}
\begin{equation}
\bar\gamma_0\,=\,\gamma_0\,+\,\frac{1}{2}\kappa H_0^2
\end{equation}
Other quantities of the order zero come from differential geometry and are essential to calculate the viscous membrane force $\pmb{f}_v$. They are provided in the appendix unless :
\begin{equation}
H^{(0)}\,=\,-\frac{1}{2R}\: ; \:K^{(0)}\,=\,0
\end{equation}
\begin{equation}
\pmb{t}^{(0)}_\theta\,=\,R\pmb{e_\theta}\: ; \:\pmb{t}^{(0)}_z\,=\,\pmb{e_z}\: ;\: \pmb{n}^{(0)}\,=\,\pmb{e_r}
\label{vec0}
\end{equation}

\section{Linear analysis of stability }
\subsection{Notation}
The perturbation of membrane shape  $\delta R(z,t)$ is expanded on normal modes of wavenumber $k$:
\begin{equation}
\delta R\,=\,\sum_k\: \delta R_k\,=\, \sum_k\: R_k\,e^{st\,+\,jkz}\:+\: cc
\end{equation}
with $\mid R_k\mid << R$, $j^2\,=\,-1$. $R_k$ is the amplitude of the perturbation of mode k. $cc$ means complex conjugate. All the physical quantities are developed in the same way:
\begin{equation}
\delta A\,=\,\sum_k\: \delta A_k\:+\: cc\,=\, \sum_k\: A^{(i,o)}_k(r)\,e^{st\,+\,jkz}\:+\: cc
\end{equation}
where $\delta A$ are the pressure $\delta p^{(i,o)}$, the radial $\delta V_r^{(i,o)}$ or longitudinal $\delta V_z^{(i,o)}$ velocities in the inner and outer volumes. The amplitude is a function of the radial coordinate $r$.

\subsection{Membrane incompressibility and viscous force}
Associating the bulk incompressibility of fluids (eq. \ref{divv}) to the membrane incompressibility (eq. \ref{divvs}) permit to simplify the second constraint to:
\begin{equation}
\big(\pmb{n}.\bar{\bar{\pmb{D}}}.\pmb{n}\big)_{\pmb{x}_m}\,=\,0
\end{equation}
Using eq. \ref{vec0}, the linearization leads to $\pmb{e_r}.\pmb{\delta}\bar{\bar{\pmb{D}}}.\pmb{e_r}\,=\,0$ and becomes in cylindrical coordinates:
\begin{equation}
\Bigg(\frac{\partial \delta V_r^{(i,o)}}{\partial r}\Bigg)_{r=R}\,=\,0
\label{eq:eqspec}
\end{equation}

In the basic state, velocity is zero (eq. \ref{v00}): $V^{\beta,(0)}=V_n^{(0)}=0$. Thus only the terms of the membrane viscous force (eq. \ref{fv}) with non zero geometrical factors in the basic state are relevant. With all the results of the appendix C, the linearized membrane viscous force is derived:
\begin{eqnarray}
\pmb{f}_v\,=\,\mu_s\Big(\frac{\partial^2 \delta V_z^{(i,o)}}{\partial z^2}-\frac{1}{R}\frac{\partial \delta V_r^{(i,o)}}{\partial z}\Big)_{r=R}\:\pmb{e_z}\,\nonumber\\
-\,\frac{2\mu_s}{R^2}\, (\delta V_r^{(i,o)})_{r=R}\: \pmb{e_r}
\end{eqnarray}
The perturbed force has still a normal component, a property of a bidimensionnal flow on a curved surface.

\subsection{Hydrodynamics}
The pressure is an harmonic function. In the cylindrical geometry,  Using the condition of existence at the center of the cylinder $r=0$ and when r tends to infinity, we derive:
\begin{equation}
\delta p_k^{i,(1)}\,=\,p^i_k\,I_0(kr)\,e^{st\,+\,jkz}
\end{equation}
\begin{equation}
\delta p_k^{o,(1)}\,=\,p^o_k\,K_0(kr)\,e^{st\,+\,jkz}
\end{equation}
where $I_0$ and $K_0$ are modified Bessel functions of first and second kind of order 0. 
The fluid velocities are calculated from Stokes equation (Eq. \ref{stokes}):
\begin{equation}
\delta V^i_{r,k}\,=\,\Big(u_k^iI_1(kr)\,+\,\frac{p^i_k}{2\eta_i}\,rI_0(kr)\Big)\,e^{st\,+\,jkz}
\end{equation}
\begin{equation}
\delta V^i_{z,k}\,=\,\Big(v_k^iI_0(kr)\,+\,\frac{jp^i_k}{2\eta_i}\,rI_1(kr)\Big)\,e^{st\,+\,jkz}
\end{equation}
\begin{equation}
\delta V^o_{r,k}\,=\,\Big(u_k^oK_1(kr)\,+\,\frac{p^o_k}{2\eta_o}\,rK_0(kr)\Big)\,e^{st\,+\,jkz}
\end{equation}
\begin{equation}
\delta V^o_{z,k}\,=\,\Big(v_k^oK_0(kr)\,-\,\frac{jp^o_k}{2\eta_o}\,rK_1(kr)\Big)\,e^{st\,+\,jkz}
\end{equation}
where $I_1$ and $K_1$ are modified Bessel functions of first and second kind of order 1. 

The fluid is incompressible (Eq. \ref{divv}) leading to two relations between inner and outer coefficients:
\begin{eqnarray}
\label{divv_vk}
p^i_k\,+\,\eta_ik(u_k^i\,+\,jv_k^i)\,=\,0\\
p^o_k\,+\,\eta_ok(-u_k^o\,+\,jv_k^o)\,=\,0
\end{eqnarray}

\subsection{Boundary conditions at membrane}

The tangent and normal velocities are continuous at the membrane (eqs. \ref{contVn}, \ref{contVt}):
\begin{eqnarray}
v_k^iI_0+\frac{jp_k^i}{2\eta_i}RI_1\,=\,v_k^oK_0\,-\,\frac{jp_k^o}{2\eta_o}RK_1\\
u_k^iI_1+\frac{p_k^i}{2\eta_i}RI_0\,=\,u_k^oK_1\,+\,\frac{p_k^o}{2\eta_o}RK_0
\label{contV_coeffk}
\end{eqnarray}
where the argument of Bessel functions is $kR$.

The new equation \ref{eq:eqspec} dealing with the membrane incompressibility allows to calculate simply two new relations:
\begin{eqnarray}
u_k^i\,=\,\frac{p_k^iR}{2\eta_i}\:\frac{I_0\,+\,kRI_1}{I_1\,-\,kRI_0}\\
u_k^o\,=\,\frac{p_k^oR}{2\eta_o}\:\frac{K_0\,-\,kRK_1}{kRK_0\,+\,K_1}
\label{divvs_uk}
\end{eqnarray}
where the relations $I'_0=I_1$, $K'_0=-K_1$, $(xI_1)'=xI_0$ and $(xK_1)'=-xK_0$ are used.
Solving all the equations \ref{divv_vk}-\ref{divvs_uk} permit to establish the relation between inner and outer perturbations of  pressures:
\begin{equation}
p_k^o\,=\,p_k^i\:\frac{\eta_o}{\eta_i}\:\frac{kRK_0+K_1}{I_1-kRI_0}\:\frac{2I_0I_1+kR(I_1^2-I_0^2)}{2K_0K_1+kR(K_0^2-K_1^2)}
\label{pko}
\end{equation}
Note that here the ratio of viscosities appears.

All the relations between the coefficients are valid whatever the forces involved in the mechanical equilibrium. The tangent one permit to determine the variation of the mechanical tension $\gamma = \gamma_0+\delta\gamma$:
\begin{eqnarray}
\frac{\partial\delta \gamma}{\partial z}+\eta_o\Big(\frac{\partial \delta V_z^o}{\partial r}+\frac{\partial \delta V_r^o}{\partial z}\Big)-\eta_i\Big(\frac{\partial \delta V_z^i}{\partial r}+\frac{\partial \delta V_r^i}{\partial z}\Big)\nonumber\\
+\mu_s\Big(\frac{\partial^2 \delta V_z^{(i,o)}}{\partial z^2}-\frac{1}{R}\frac{\partial \delta V_r^{(i,o)}}{\partial z}\Big)=0
\end{eqnarray}
with the equation applied at $r=R$.
To understand the different contributions to the tension, we first investigate the case $\mu_s=0$:
\begin{eqnarray}
\gamma_k^{\mu_s=0}=\frac{p_k^o}{k}K_1+\frac{p_k^i}{k}I_1-p_k^oR\Big(K_0+K_1\frac{K_0-kRK_1}{kRK_0+K_1}\Big)\nonumber\\+p_k^iR\Big(I_0+I_1\frac{I_0+kRI_1}{I_1-kRI_0}\Big)
\end{eqnarray}
The third and fourth terms cancel without contrast of viscosity between inside and outside (see \ref{pko}). In the case $\mu_s \neq 0$, the mechanical tension is: 
\begin{eqnarray}
\gamma_k=\gamma_k^{\mu_s=0}+\frac{\mu_s}{\eta_i}\,p_ki\,\frac{2I_0I_1+kR(I_1^2-I_0^2)}{I_1-kRI_0}
\end{eqnarray}
Consider now the normal mechanical equilibrium taking anti account the membrane incompressibility constraint (eq. \ref{eq:eqspec}):
\begin{eqnarray}
\delta p^i-\delta p^o-\frac{\delta\gamma}{R}-\frac{2\mu_s}{R^2}\, \delta V_r^i\nonumber\\
=\,\kappa(2\frac{\partial^2\delta H}{\partial z^2}+\frac{3}{R^2}\delta H+\frac{2}{R}\delta K)\nonumber\\
+2\kappa H_0\delta K \,-2\bar\gamma_0\delta H
\label{normalequilin}
\end{eqnarray}
where $\delta H = H-H^{(0)}=H+1/2R$ and $\delta K=K-K^{(0)}=K$. Their expressions are provided in the appendix D as a function of $\delta R$ and its derivatives. This equation has been separated in two members, the right member which depend on $\delta H$ and $\delta K$ and thus on $\delta R$ and the left member with terms of pressure, membrane viscous force and tension which depend on the coefficients $p_k^{(i,o)}$ after little algebra. The right member will provide the criteria of instability while the left one contributes to the dynamics, the characteristic time of growing perturbation. Eq. \ref{normalequilin} provides the relation between the amplitudes of pressure $p_k^i$ and and shape perturbation $R_k^{(1)}$.

%%%%%%%%%%%%%%%%%%%%%%%%%%%%%%%%%%%%%%%%%%%%%%%%%%%%%%
%%%%%%%%%%%%%%%%%%%%%%%%%%%%%%%%%%%%%%%%%%%%%%%%%%%%%%
%%%%%%%%%%%%%%%%%%%%%%%%%%%%%%%%%%%%%%%%%%%%%%%%%%%%%%
%%%%%%%%%%%%%%%%%%%%%%%%%%%%%%%%%%%%%%%%%%%%%%%%%%%%%%
%[htbp]

%%%%%%%%%%%%%%%%%%%%%%%%%%%%%%%%%%%%%%%%%%%%%%%
\section{Results and discussion}
\subsection{dispersion relation in the general case}
To obtain the characteristic time of growth or relaxation of a shape perturbation $\delta R$, the continuity of the normal membrane velocity satisfies with eqs:
\begin{equation}
\frac{\partial \delta R_k}{\partial t}\,=\, s \delta R_k\,=\,\delta V_{r,k}^{(i,o)}
\end{equation}
leading to the following relation between the amplitudes :
\begin{equation}
s R_k^{(1)}\,=\,u_k^iI_1\,+\,\frac{p_k^i}{2\eta_i}RI_0
\end{equation}
Using , the dispersion relation $s = s(k)$ is provided by the following general equation what is the original analytical result of this paper:
\begin{equation}
s(k)\, = \, -\frac{\kappa}{2R^3}\frac{Q(k)}{(1\,+\,k^2R^2)\,D(k)}
\end{equation}
\begin{equation}
Q(k)\,=\,kR\,(R^4k^4\,+\,bR^2k^2\,+\,c)
\end{equation}
\begin{eqnarray}
D(k)\,=\,\frac{\mu_s}{R}\,\frac{kR}{1\,+\,k^2R^2}\,+\,\eta_i\frac{I_1^2}{2I_0I_1+kR(I_1^2-I_0^2)}\nonumber \\
+\,\eta_o\frac{K_1^2}{2K_0K_1+kR(K_0^2-K_1^2)}
\end{eqnarray}
associated with the following definitions of the constants:
\begin{eqnarray}
b\,=\,\frac{\bar\gamma_0 R^2}{\kappa}\,-\frac{1}{2}\,+\,2H_0R\nonumber\\=\frac{\gamma_0 R^2}{\kappa}\,-\frac{1}{2}\,+\,2H_0R+\frac{1}{2}H_0^2R^2
\end{eqnarray}
\begin{eqnarray}
c\,=\,\frac{3}{2}\,-\,\frac{\bar\gamma_0 R^2}{\kappa}\nonumber\\
= \,\frac{3}{2}\,-\,\frac{\gamma_0 R^2}{\kappa}-\frac{1}{2}H_0^2R^2
\end{eqnarray}

The polynome $P(k)$ can be determined by the minimization of the energy of a modulated lipidic tube and has been performed by several authors \cite{Granek1995JPhys,Chaieb1998PRE}. The product $(1+k^2R^2)D(k)$ is often called the dynamical factor which characterizes how fast the perturbation is damped or amplified. It takes into account the hydrodynamic dissipation in the outer and inner bulk and also here, the hydrodynamic dissipation along the membrane. Sometimes, it was proposed that this term is the same than the well-known Tomotika result \cite{Granek1996Langmuir} as if the lipidic or copolymer membrane has the same dynamical behavior than a fluid-fluid interface governed by surface tension. This point of view neglected in fact the incompressible nature of a lipidic or copolymer membrane and the gradient of the mechanical tension due to this constraint. However, a more accurate analysis was undertaken notably to understand the laser-induced pearling of lipidic tethers \cite{BarZiv1994PRL,BarZiv1998BiophysJ} providing a new dynamical factor \cite{Nelson1995PRL,Gurin1996JETP,Goldstein1996JPhys,Powers1997PRL,Powers2010RMP}. Nevertheless, it has been already shown \cite{Boedec2014JFM} that these results were not in agreement with previous results on the stability of a highly viscous cylinder fluid with strong surfactants \cite{Timmermans2002JFM}. Finally, if the shear membrane viscosity and the spontaneous curvature are set to zero, we recover the same expressions than two previous recent studies \cite{Boedec2014JFM,Narsimhan2014JFM}. \\
To our knowledge, the unique expression which takes into account the membrane dissipation has been reported in the previous work \cite{Powers2010RMP} which does not consider the spontaneous curvature. However, as written before,  the tension gradients was neglected leading to. a different expression than ours.. 

%In the case of a cylindrical copolymer micelle, there is no solvent inside. A minimal model considers a single curved membrane, a cylindrical micelle embedded in an external solvent. This model is crude but exhibits the salient ingredients: curvature and tension effects with a dynamics controlled by viscous dissipation. The calculation is simpler than previously considering only membrane and external viscosity and can be deduced here just by omitting the internal contribution:
%\begin{equation}
%s(k)\, = \, -\frac{\kappa}{2R^3}\frac{kR\,(R^4k^4\,+\,bR^2k^2\,+\,c)}{\frac{\mu_s}{R}\,kR+\eta_o\frac{K_1^2(1\,+\,k^2R^2)}{2K_0K_1+kR(K_0^2-K_1^2)}}
%\end{equation}
%with the same expressions for the coefficients $b$ and $c$. However, it is noteworthy that the physical parameters, notably the bending modulus, the tension and the spontaneous curvature can be different. 

\begin{figure}[htbp]
\begin{center}
\includegraphics[scale=0.31]{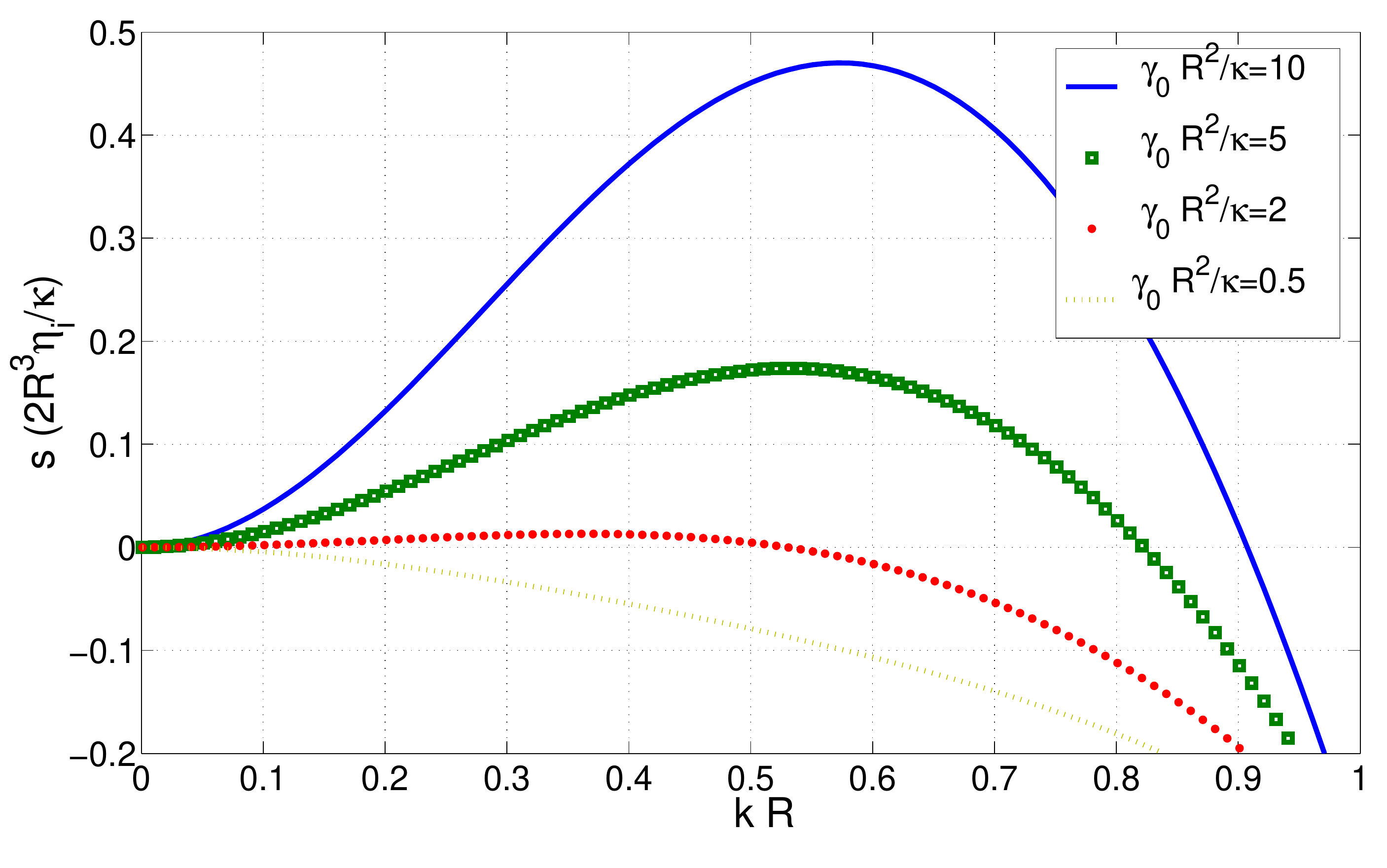}
\caption{Variation of the dispersion relation with the mechanical tension $\gamma_0$. The mean curvature $H_0$ and the surface viscosity $\mu_s$ are set to zero. The growth rate s is made dimensionless with the characteristic time $2R^3/(\kappa\eta_i)$ based on the inner bulk viscosity. The membrane contrast is equal to one, it means the same internal and external bulk viscosities. The tether is stable for $\gamma_0R^2/\kappa=0.5$ and unstable for $\gamma_0R^2/\kappa=2,\, 5,\, 10.$}
\label{fig:fig1}
\end{center}
\end{figure}

\subsection{the case without spontaneous curvature and shear membrane viscosity}
\label{tensioninduced}
Here, we recalled briefly the results when only the membrane tension plays a role in agreement with two previous studies \cite{Boedec2014JFM,Narsimhan2014JFM} to make clear the contribution of spontaneous curvature and shear membrane viscosity in the following. Thus, we set: $H_0 = 0$ and $\mu_s = 0$.\\
If we only keep the pressure and the membrane terms that do not depend on space, the pressure perturbation can be written using the normal equilibrium (eq. \ref{normalequilin}) and appendix D:
\begin{equation}
\delta p^i - \delta p^o \approx \frac{\delta R}{R^2}\Big(\frac{3\kappa}{2R^2}-\gamma_0\Big)
\end{equation}
Where the radius is increased (decreased) by the perturbation $\delta R$, the pressure diminishes (raises) if the membrane tension $\gamma_0$ exceeds a critical value $\gamma_c$, a function of the bending elastic modulus and the radius: $\gamma_0 > \gamma_c$ with $\gamma_c = 3 \kappa/ 2R^2$. This corresponds to $b > 0$ and $c < 0$. In this case, a longitudinal flow amplifies the initial perturbation: the membrane tube is unstable. This result is known and has also been derived by energetic analysis. It highlights the contribution of bending energy. Indeed, for a fluid-fluid interface, the critical surface tension of the Rayleigh-Plateau instability is zero and not a finite value as here. At equilibrium, the membrane tension is $\gamma = \kappa/2R^2$ obtained by the minimization of the Helfrich and tension energies for a cylinder. It means that an external force must be applied to reach the threshold $\gamma_c$ with a laser, an electric field, a flow such as in elongationnal configuration or sedimentation. If a force $f$ pulls the tether at each tip, the critical force $f_c$ is given by: $f_c = (\gamma_c+\frac{\kappa}{2R^2})2\pi R =  (8/3)\pi R\gamma_c$, an expression different from  $2\pi R \gamma_c$ given for a fluid-fluid interface which still underlines the role of bending energy. Another characteristic of this instability is that all the wavenumbers in the range $[0; k_0]$ are unstable as their growth rate $s(k)$ is positive (see figure \ref{fig:fig1}):
\begin{equation}
\label{eq:k0}
k_0=\frac{1}{R}\Bigg(\frac{1}{4}-\frac{\gamma_0R^2}{2\kappa}+\frac{1}{4}\Big(4(\frac{\gamma_0R^2}{\kappa})^2+12\frac{\gamma_0R^2}{\kappa}-23\Big)^{1/2}\Bigg)^{1/2}
\end{equation}
One of the unstable modes grows faster and is called the most unstable wavenumber $k_m$ that increases with the dimensionless  membrane tension $\gamma_0R^2/\kappa$ and with the ratio of viscosities $\eta_i/\eta_o$. 

With the same viscosities inside and outside $\eta_i=\eta_o$, $k_m \approx 0.56/R$ what corresponds to a growth rate $s_m = s(k_m) \approx 0.24 \kappa/\eta R^3$. The characteristic time is given by the competition between bending resistance and viscous stress. Consider the following parameter to have an order of magnitude:  a radius $R = 1 \mu$m, a lipidic bending modulus $\kappa = 20 k_BT$ and the water viscosity $\eta = 1$ m.Pa.s: the wavelength $\lambda_m=k_m/2\pi\approx 11 \mu$m and $1/s_m \approx 50$ ms. Relevant values of the difference of viscosities between inside and outside and membrane tension leads to a 10 $\%$ variation of $k_m$.  A complete analysis of variations of $k_m$ and $s_m$ with the membrane tension and the bulk viscosities can be found in our previous study \cite{Boedec2014JFM}. 

\subsection{the case with spontaneous curvature and without shear membrane viscosity}
\begin{figure}[htbp]
\begin{center}
\includegraphics[scale=0.34]{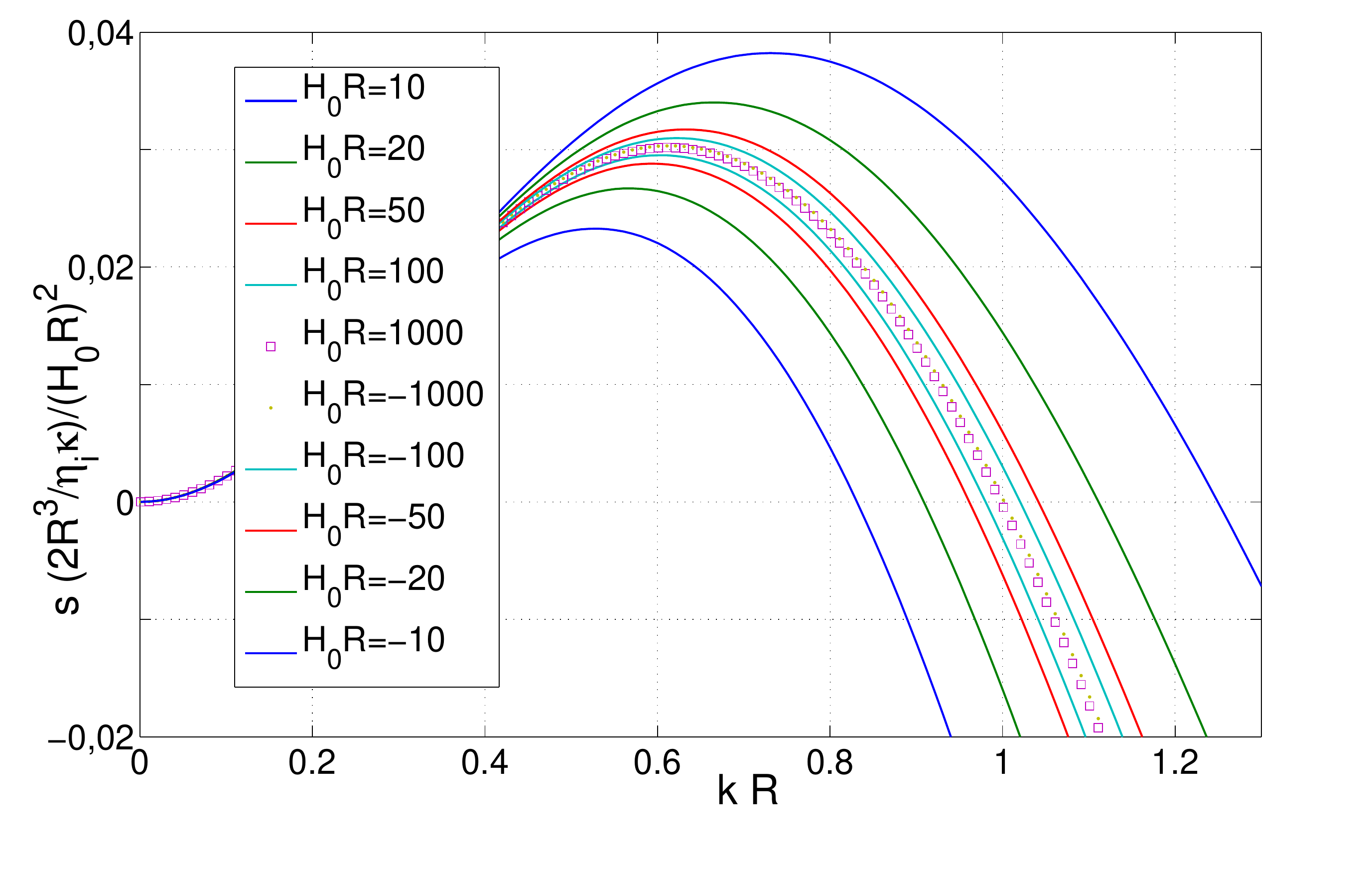}
\caption{Variation of the dispersion relation with the spontaneous curvature $H_0$ in the regime of high $H_0$. The mechanical tension $\gamma_0$ and the membrane viscosity $\mu_s$ are set to zero, The dispersion relation s(k) tends to the curve $s(k) \approx \eta_i\kappa H_0^2/(2R)$ in the limit of high spontaneous curvature. The wavenumber which cancels the growth rate is $1/R$ and the wavenumber $k_m$ of maximal growth rate is approximately $0.61 / R$ for large $H_0$. These expressions can provide good orders of magnitude in a large range of $H_0$ in a first approximation.}
\label{fig:fig2}
\end{center}
\end{figure}

\begin{figure}[htbp]
\begin{center}
\includegraphics[scale=0.33]{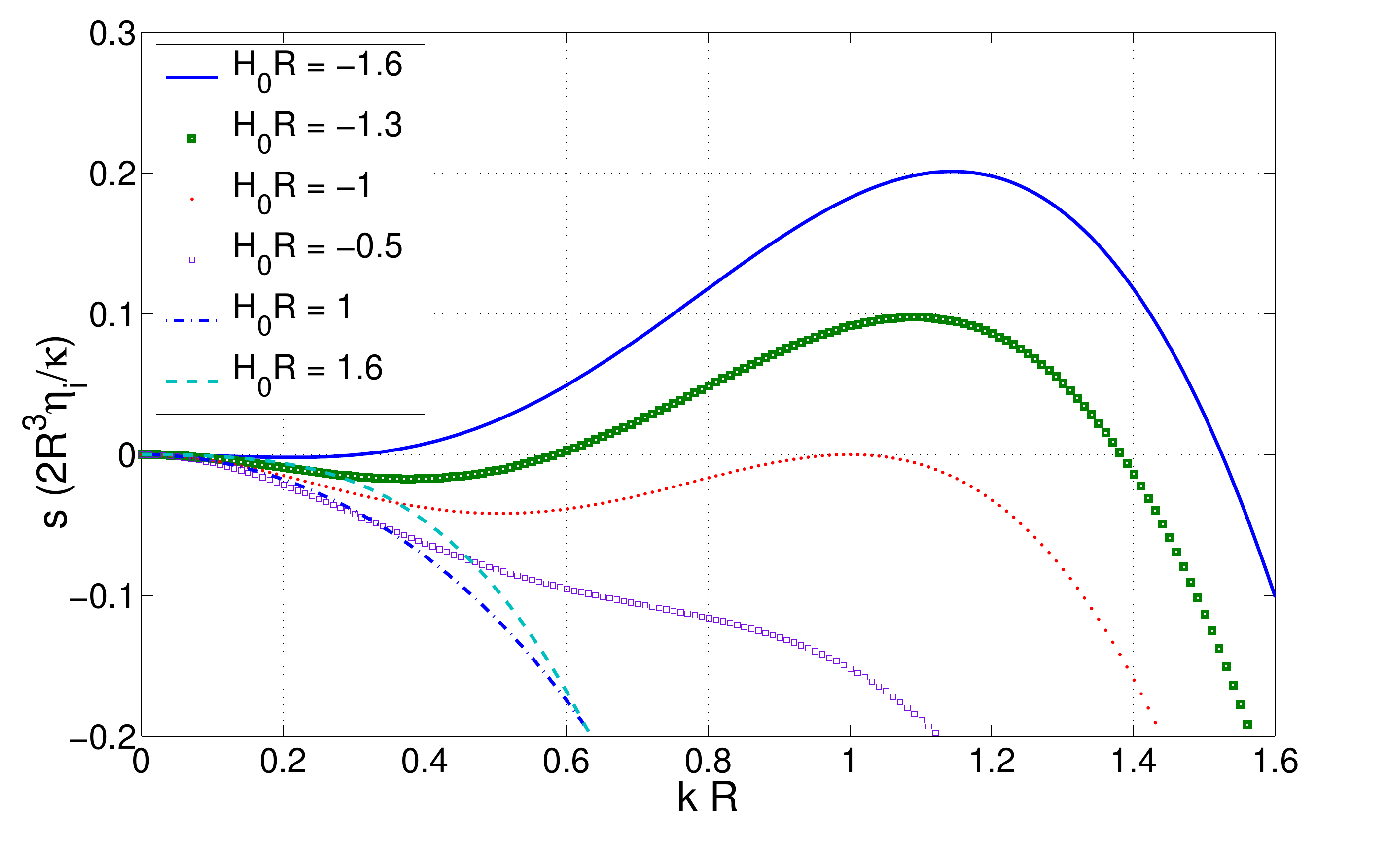}
\caption{Variation of the dispersion relation with the spontaneous curvature $H_0$ in the regime of moderate $H_0$ range. The mechanical tension $\gamma_0$ and the membrane viscosity $\mu_s$ are set to zero, the internal and external bulk viscosities are equal. Note that with our convention, the curvature of a sphere is negative, a classic way in modeling of vesicles under flow. The tether is unstable for $H_0 = -1.6/R,\,-1.3/R$ and stable for $H_0 = -0.5/R,\,1/R,\,1.6/R$. The transition is at $H_0=-1/R$. The negative and positive values of $H_0$ have been chosen such that $(1/2)H_0^2R^2 < 3/2$, the threshold of the instability driven by the mechanical tension.}
\label{fig:fig3}
\end{center}
\end{figure}

We recall the convention on the sign of $H_0$: the curvature of a sphere and by extension of a cylinder is negative. It comes from a preliminary  choice using differential geometry to describe more easily the variations of the geometrical properties along the surface. To come back to the other convention (a positive curvature for a sphere or a cylinder), $H_0$ has to be replaced by $- H_0$. Then, the only difference is the constant $b$ which becomes $b\,=\,\frac{\bar\gamma_0 R^2}{\kappa}\,-\frac{1}{2}\,-\,2H_0R$. 

\begin{figure}[htbp]
\begin{center}
\includegraphics[scale=0.32]{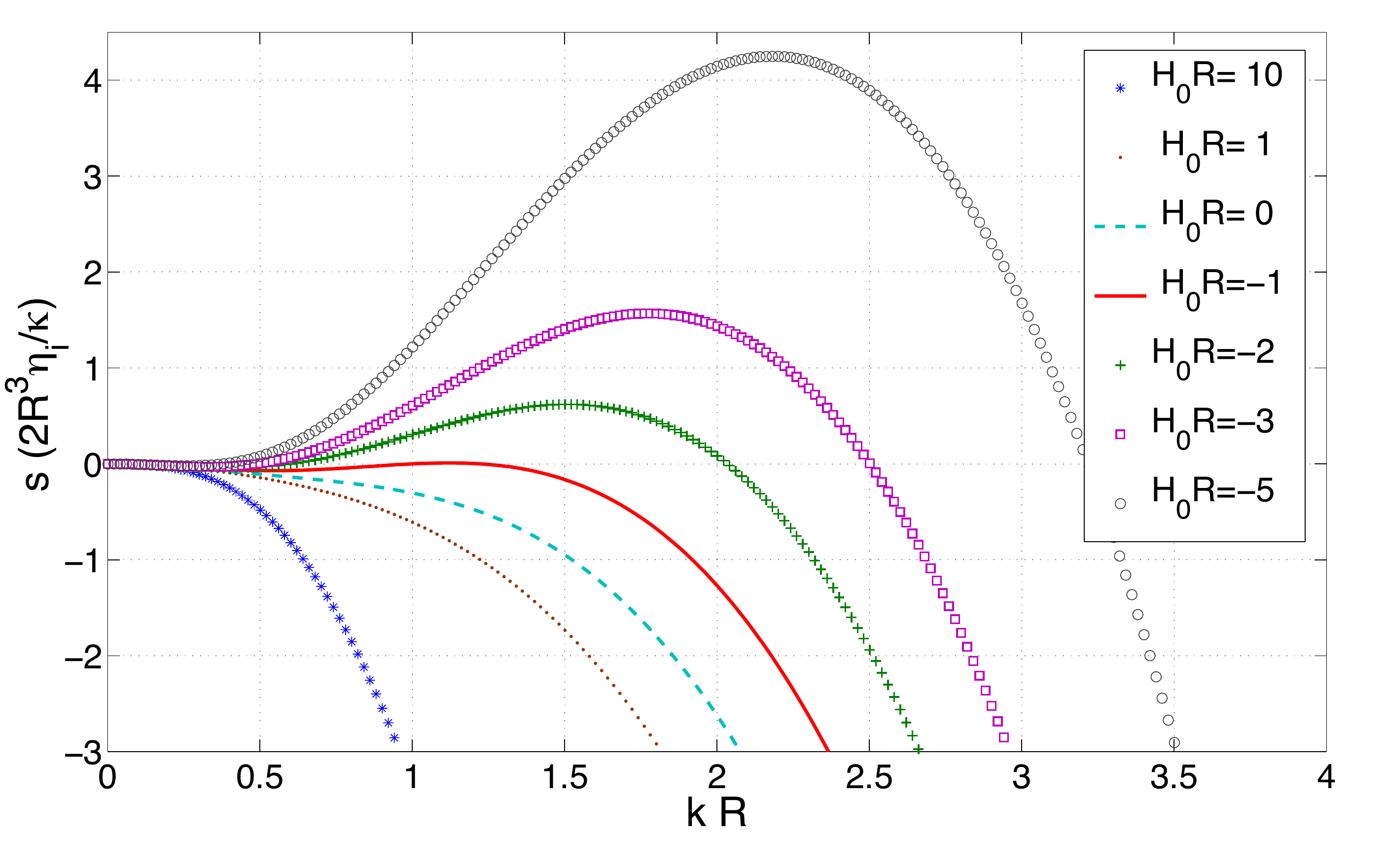}
\caption{Variation of the dispersion relation with the spontaneous curvature $H_0$ in the case of a zero effective membrane tension $\bar{\gamma}_0 = 0$.}
\label{fig:fig4}
\end{center}
\end{figure}

The spontaneous curvature $H_0$ appears either in the effective membrane tension $\bar\gamma_0$ or by the variation of the gaussian curvature (see eq. \ref{equimec}) which provides a linear variation of the coefficient b with $H_0$. This linear term disappears if the membrane is flat (see eq. \ref{deltaK}) highlighting the difference with  previous studies \cite{Shi2014ACIS}. Investigating the signs of b and c makes clear that at large spontaneous curvature, the coefficients are dominated by the square of $H_0$ while for weak values, the linear term is essential. For the sake of simplicity, we only investigate the limits of zero tension $\gamma_0 = 0$ (figures \ref{fig:fig2} and \ref{fig:fig3}) and zero effective tension $\bar\gamma_0 = 0$ (figure \ref{fig:fig4}). This choice has a degree of  which provide a clear picture of the contribution of spontaneous curvature.

In the case without mechanical tension ($\gamma_0=0$), the limits of high (figure \ref{fig:fig2})  and small to moderate (figure \ref{fig:fig3})  spontaneous curvature present a different relation of dispersion, i.e the variation of the growth rate $s$ of a perturbation in respect of its wavenumber $k$. Indeed, in the limit of large values of the spontaneous curvature, the cylinder membrane is always unstable due to the effective membrane tension $\bar\gamma_0 = \kappa H_0^2/2 >> 3/2$ explaining shape instability ($b>0$ and $c<0$): see figure \ref{fig:fig2}. As expected, the dispersion relation is analog to the tension-induced pearling instability which is recalled in the section \ref{tensioninduced}. The marginal mode is the zero wavenumber (figure \ref{fig:fig2}). In the case of small to moderate spontaneous curvature (figure \ref{fig:fig3}), the tether becomes unstable above the threshold $H_{Oc}=-1/R$ corresponding  to the marginal mode $k_c=1/R$. A finite range of wavenumbers $[k_1;k_2]$ with $k_1>0$ is unstable contrary to tension-induced pearling. For $H_0 R = -1.8$, the most unstable mode $k_c=2\pi/\lambda_c$ leading to  a wavelength $\lambda_c \approx 5.38 R$. Here, the term $2\kappa H_0\delta K \approx 2\kappa H_0k^2\delta R / R$ of the normal mechanical equilibrium eq. \ref{normalequilin} varies linearly with $H_0$ and governs the amplified perturbation. A necessary condition is a negative spontaneous curvature: $H_0 < 0$ corresponding to the same sign of the tether's curvature. This is strikingly different from the effective tension-induced instability where $H_0$ and $-H_0$ play the same role. Thus, this mode of instability is called spontaneous curvature-induced instability. 

In the case without effective mechanical tension ($\bar\gamma_0=0$), the tether is unstable if $H_0 R < 1/4-\sqrt{3/2} \approx -0.97$. A negative spontaneous curvature is a prerequisite. A finite range of wavenumbers $[k_1,k_2]$ with $k_1>0$ is unstable meaning that the instability is the same as the spontaneous curvature-induced pearling instability. In the limit of high $-H_0 R$, $k_1 \approx -3/4H_0R^2$ explaining why this statement is not clear in  figure \ref{fig:fig4}. The most unstable mode has a higher wavenumber and a higher growth rate than the case without mechanical tension. 

\subsection{the cases with shear membrane viscosity}

\begin{figure}[htbp]
\begin{center}
\includegraphics[scale=0.32]{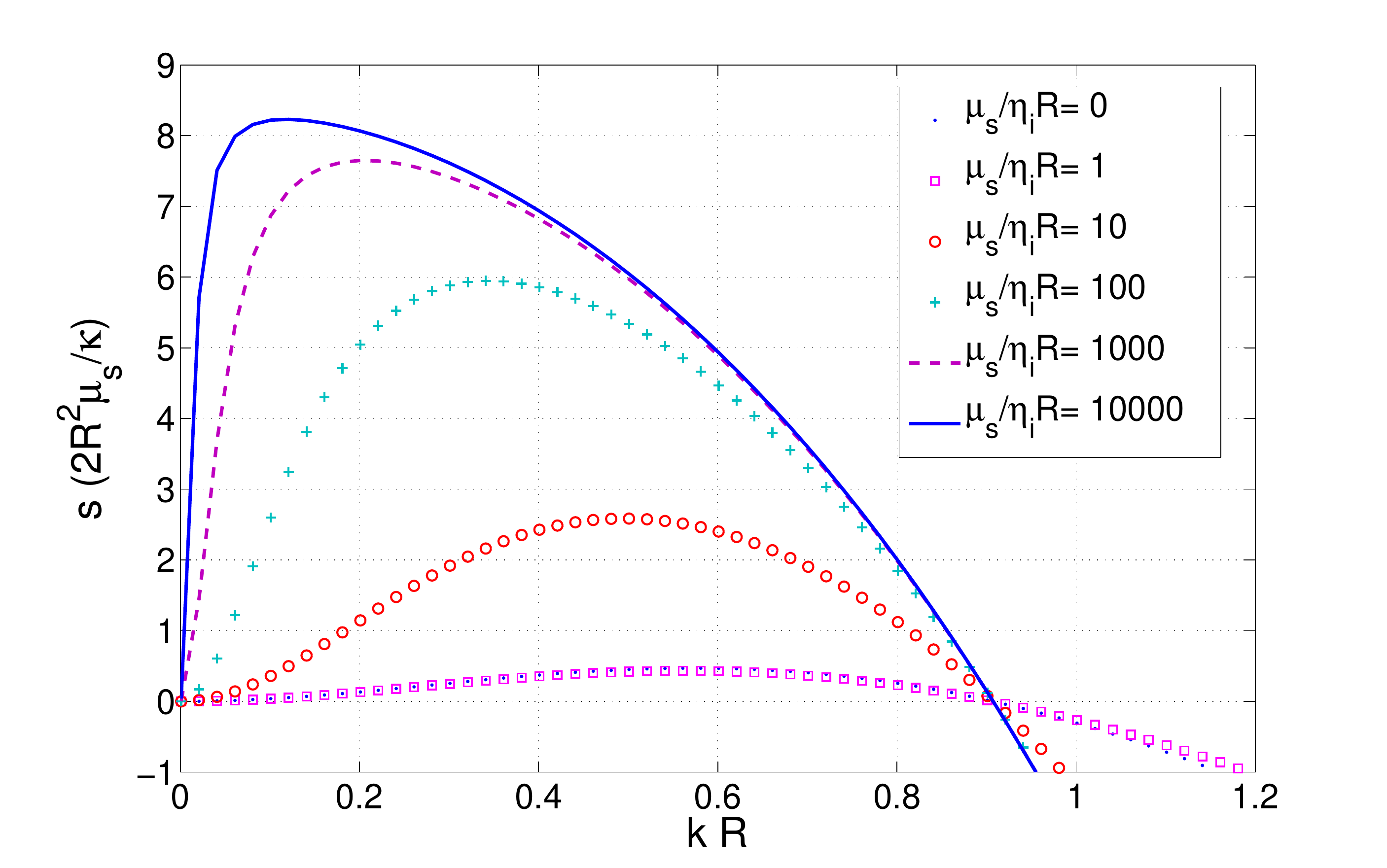}
\caption{Variation of the dispersion relation $s(k)$ with the membrane viscosity $\mu_s$ in the case of a zero spontaneous curvature $H_0 = 0$. Note that here, the growth rate $s$ is made dimensionless using the membrane viscosity contrary to the previous figures. The dimensionless membrane tension is set to  $R^2\gamma_0/\kappa = 10$. The internal and external bulk viscosities are equal. The membrane viscosity has a stronger effect in the linear selection of the wavelength of the pattern than bulk viscosities.}
\label{fig:fig5}
\end{center}
\end{figure}

\begin{figure}[htbp]
\begin{center}
\includegraphics[scale=0.33]{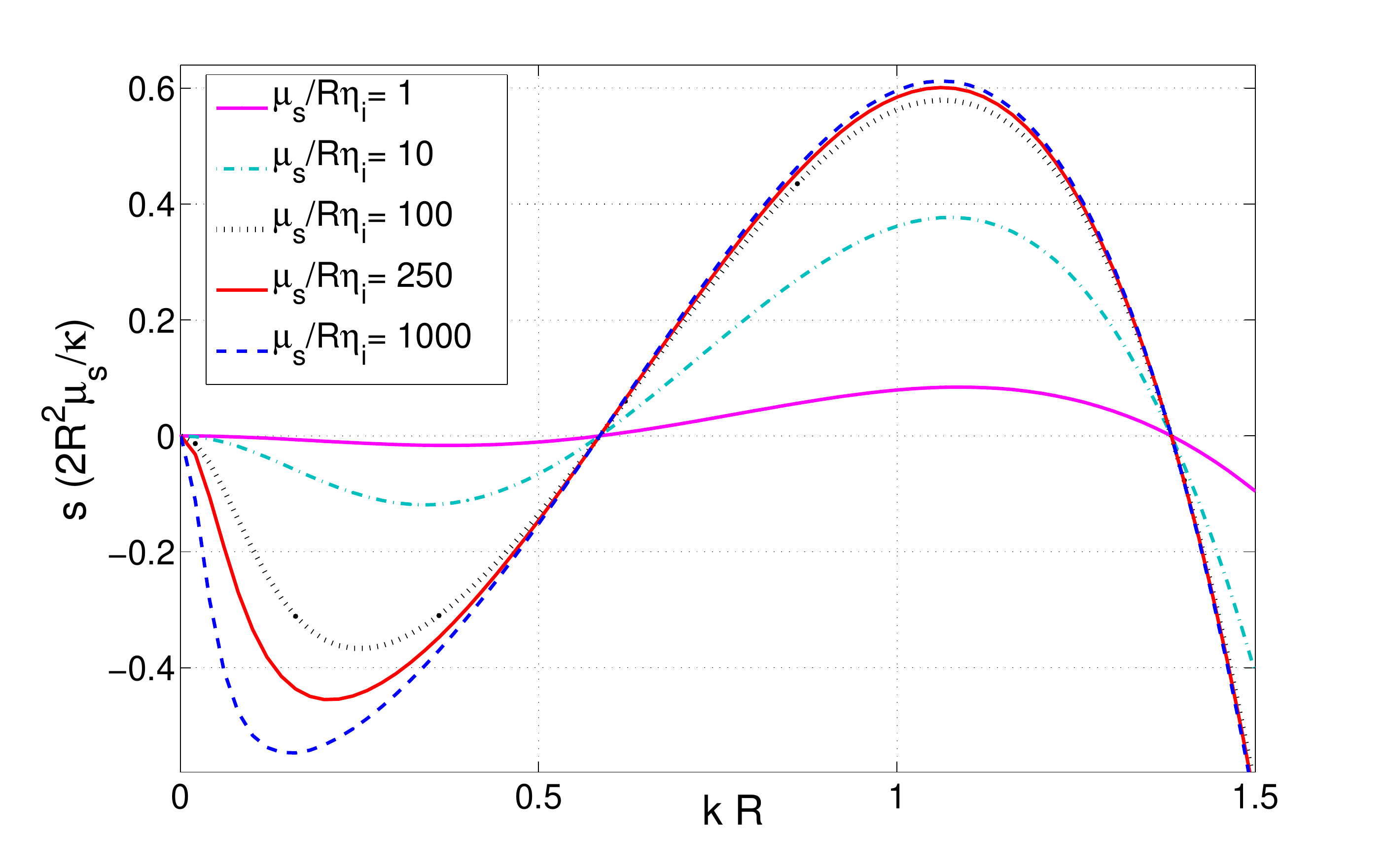}
\caption{Variation of the dispersion relation $s(k)$ with the membrane viscosity $\mu_s$ in the case of a zero membrane tension $\gamma_0 = 0$. Note that here, the growth rate $s$ is made dimensionless using the membrane viscosity. spontaneous curvature is set to $H_0 R = -1.3$. The internal and external bulk viscosities are equal. Contrary to the previous case of instability driven by membrane tension, the membrane viscosity has not a minute effect in the linear selection of the wavelength of the pattern. In the limit of high membrane viscosity, the most unstable mode $k_m$ is unique characterized by $k_m \approx 1.06 / R$ and $1/s_m \approx 1.22 R^2 \mu_s/\kappa$.}
\label{fig:fig6}
\end{center}
\end{figure}

Previously, the only dissipation included in the modeling was the one due to the long range hydrodynamical flow around and inside the tether coupled with the shape perturbation. In this part, the shear membrane viscosity $\mu_s$ is introduced. The aim is to determine its role in the characteristic time of the process  and the selected characteristic wavelength $\lambda_m=2\pi/k_m$. We expect at least a strong slowing down of any dynamics. This is demonstrated in  figures \ref{fig:fig5} and \ref{fig:fig6} to compare to figures \ref{fig:fig1}-\ref{fig:fig4}.\\
Two modes have been previously identified with two different typical dispersion relations $s(k)$: a membrane tension-induced instability (or effective membrane tension) and a specific spontaneous curvature-induced instability.\\
First, consider the case without spontaneous curvature: $H_0 = 0$. As for bulk dissipation, the shear membrane viscosity appears  only in the dynamical factor $(1+k^2R^2)D(k)$ which contributes to select the most unstable mode $k_m$ and the characteristic time $1/s(k_m)$ associated with this mode. Thus, the threshold is always given by the critical membrane tension: $\gamma_0 > \gamma_c$ with $\gamma_c=\frac{3\kappa}{2R^2}$. The dispersion relation has a typical variation with a large range of unstable wavenumbers $[0; k_0]$ with two marginal modes 0 and $k_0$ given by the relation \ref{eq:k0}. Note that $k_0$ does not depend on the membrane and bulk viscosities. Whatever the membrane tension above the critical one, the most unstable wavenumber $k_m$ tends to zero with increasing the membrane viscosity: figure \ref{fig:fig5}. Indeed, if the following parameters are considered $\gamma_0R^2/\eta_i=10$, $k_m$ varies from $\frac{0.56}{R}$ m$^{-1}$ for $\mu_s=0$ to $k_m = \frac{0.12}{R}$ m$^{-1}$ for $\mu_s/\eta_iR=10^4$. For example, with $R = 1 \mu$m, $\lambda_m \approx 11\,\mu$m and $\lambda_m \approx 52\,\mu$m respectively. The membrane tension-induced shape instability promotes large wavelengths compared to the radius, a high shear membrane viscosity amplifying this effect. 

%%%%%%%%%%%%%%%%%%%%%%%%%%%%%%%%%%%%%%%%%%%%%%%%%%%%%%
%%%%%%%%%%%%%%%%%%%%%%%%%%%%%%%%%%%%%%%%%%%%%%%%%%%%%%
%%%%%%%%%%%%%%%%%%%%%%%%%%%%%%%%%%%%%%%%%%%%%%%%%%%%%%
%%%%%%%%%%%%%%%%%%%%%%%%%%%%%%%%%%%%%%%%%%%%%%%%%%%%%%

\section{Conclusion}
 
In polymersomes, membrane dissipation prevails over bulk dissipation in most configurations. Indeed, numerical evaluations show that the Saffman-Delbr\^{u}ck length $L_{sd}$ is larger than the typical size of a polymersome. All the results and discussions might be presented consider the Boussinesq number $B_{q_{s}}=\eta_{s_{s}}/\eta R = L_{sd} / R$ which is less than the unity in polymersomes. Without spontaneous curvature, we revovered the results of literature. With spontaneous curvature, the dispersion relation is established. With spontaneous curvature and shear surface viscosity, there is a novel instability with its own dispersion relation (figure \ref{fig:fig6}) which permits to select a characteristic size (most unstable mode) associated with a characteristic time governed by membrane dissipation only. Indeed, only a finite range of wavenumbers is unstable with the minimal one which is different from zero. This result allows the determination of the shear surface viscosity by the analysis of growing modes along a tether made of copolymers.

%%%%%%%%%%%%%%%%%%%%%%%%%%%%%%%%%%%%%%%%%%%%%%%%%%%%%%
%%%%%%%%%%%%%%%%%%%%%%%%%%%%%%%%%%%%%%%%%%%%%%%%%%%%%%
%%%%%%%%%%%%%%%%%%%%%%%%%%%%%%%%%%%%%%%%%%%%%%%%%%%%%%
%%%%%%%%%%%%%%%%%%%%%%%%%%%%%%%%%%%%%%%%%%%%%%%%%%%%%%

\section*{Acknowledgements}
This work has been carried out in the framework of the ANR 2DVisc (ANR-18-CE06-0008-01). This work has benefited from the financial support of CNES and Labex  Tec21 (ANR-11-LABX-0030).

\section*{Appendix A - some elements of differential geometry}

Here, we recall some elements of differential geometry which are necessary to calculate the tension force, the bending force and especially the viscous membrane force (see Eq. \ref{fv}) which is a complex function of geometrical quantities and membrane velocity. A point of the membrane is localized by a parametrization ($s^1; s^2$):
\begin{equation}
\pmb{x}_m\,=\pmb{x}_m(s^1,s^2)
\end{equation}
The tangent vectors to the surface at $\pmb{x}_m$ are defined by:
\begin{equation}
\pmb{t}_\beta\,=\,\frac{\partial \pmb{x}_m}{\partial s^\beta}
\end{equation}
The normal unit vector at the same point is determined as usual:
\begin{equation}
\pmb{n}\,=\,\frac{\pmb{t}_1\wedge\pmb{t}_2}{\mid\mid\pmb{t}_1\wedge\pmb{t}_2\mid\mid}
\end{equation}
The contravariant basis $(\pmb{t}_1\,;\,\pmb{t}_2\,;\,\pmb{n})$ permit to define all the physical quantities at the surface. Indeed, the velocity is given by its contravariant coordinate and the normal one: see eq. \ref{Vbeta}. A dual basis called covariant $(\pmb{t}^1\,;\,\pmb{t}^2\,;\,\pmb{n})$ can be defined as:
\begin{equation}
\pmb{t}_\beta.\pmb{t}^\alpha\,=\,\delta_\beta^\alpha
\end{equation}
where $\delta^\alpha_\tau$ is the Kronecker symbol. The metric $(g)$ and curvature ($K$) tensors are given by:
\begin{equation}
g_{\alpha\beta}\,=\,\pmb{t}_\alpha.\pmb{t}_\beta\; ; \; g^{\alpha\beta}\,=\,\pmb{t}^\alpha.\pmb{t}^\beta
\end{equation}
\begin{equation}
K_{\alpha\beta}\,=\,-\pmb{t}_\alpha.\frac{\partial \pmb{n}}{\partial s^\beta}
\end{equation}
\begin{equation}
g^{\alpha\beta}\,g_{\beta\tau}\,=\,\delta^\alpha_\tau \; ; \; g\,=\,det(g_{\alpha\beta})
\end{equation}
The two metric tensors are useful to go down and up an index of tensors: $g^{\alpha\beta}A_\beta\,=\,A^\alpha$ and $g_{\alpha\beta}A^\beta\,=\,A_\alpha$. To calculate the membrane viscous force, it is necessary to calculate $g^{\alpha\tau}K_{\tau\beta}\,=\,K^\alpha_\beta$ and $g^{\beta\tau}K^\alpha_{\tau}\,=\,K^{\alpha\beta}$.

The two invariants of the curvature tensor are the mean curvature $H$ and the gaussian curvature $K$:
\begin{equation}
H\,=\,\frac{1}{2}\,K_\alpha^\alpha\,=\,\frac{1}{2}\,g^{\alpha\beta}K_{\beta\alpha}
\end{equation}
\begin{equation}
K\,=\,det(K_{\alpha\beta})
\end{equation}

Differential operators are necessary to calculate the gradient of the mechanical tension in $\pmb{f}_\gamma$ and the Laplace-Beltrami of the mean curvature of the bending force $\pmb{f}_\kappa$:
\begin{equation}
\pmb{\nabla}_s f\,=\, \Big(\frac{\partial f}{\partial s^\alpha}\Big)\,\pmb{t}^\alpha
\end{equation}
\begin{equation}
\Delta_s f\,=\, \frac{1}{\sqrt{g}}\frac{\partial}{\partial s^\beta}\Big(\sqrt{g}\,g^{\beta\alpha}\frac{\partial f}{\partial s^\alpha}\Big)
\end{equation}

%%%%%%%%%%%%%%%%%%%%%%%%%%%%%%%%%%%%%%%%%%%%%%
\section*{Appendix B - Axisymmetrical expressions in the parametrization $(\theta,\,z)$}

As explained in the section \ref{problem}, the characteristics of the studied system are well described by a local basis based on the parametrization $(\theta,\,z)$. As we consider axisymmetrical deformations, the shape of the membrane is only a function of z: 
\begin{equation}
\pmb{x}_m\,=\,f(z,t)\,\pmb{e_r}\,+\,z\,\pmb{e_z}
\label{deff}
\end{equation}
Thus, the contravariant basis $(\pmb{t}_\theta;\,\pmb{t}_z;\,\pmb{n})$ satisfies using eq. \ref{deff}:
\begin{eqnarray}
\pmb{t}_\theta\,=\,\frac{\partial \pmb{x}_m}{\partial \theta}\,=\,f\,\pmb{e_\theta}\\
\pmb{t}_z\,=\,\frac{\partial \pmb{x}_m}{\partial z}\,=\,f' \pmb{e_r}\,+\,\pmb{e_z}\\
\pmb{n}\,=\,\frac{\pmb{t}_\theta\wedge\pmb{t}_z}{\mid\mid\pmb{t}_\theta\wedge\pmb{t}_z\mid\mid}\,=\,\frac{\pmb{e_r}\,-\,f'\,\pmb{e_z}}{\sqrt{1+ f'^2}}
\end{eqnarray}
where $f'\,=\,(\partial f/\partial z)$. The metric tensor in the covariant and contravariant basis are deduced:
\begin{eqnarray}
g_{\theta\theta}\,=\,f^2\\
g_{\theta z}\,=\,g_{z\theta}\,=\,0\\
g_{zz}\,=\,1\,+\,f'^2\\
g^{\theta\theta}\,=\,\frac{1}{f^2}\\
g^{\theta z}\,=\,g^{z\theta}\,=\,0\\
g^{zz}\,=\,\frac{1}{1\,+\,f'^2}
\end{eqnarray}
The curvature tensor $K_{\alpha\beta}$ is:
\begin{eqnarray}
K_{\theta\theta}\,=\,-\frac{f}{\sqrt{1\,+\, f'^2}}\\
K_{\theta z}\,=\,K_{z\theta}\,=\,0\\
K_{zz}\,=\,\frac{f''}{\sqrt{1\,+\,f'^2}}
\end{eqnarray}
Some intermediates are necessary:
\begin{eqnarray}
K_{\theta}^\theta\,=\,-\frac{1}{f\sqrt{1\,+\,f'^2}}\\
K_{\theta}^z\,=\,K_{z}^\theta\,=\,0\\
K_{z}^z\,=\,\frac{f''}{(1\,+\,f'^2)^{3/2}}
\end{eqnarray}
Thus, the mean $H$ and gaussian $K$ curvatures are determined:
\begin{equation}
2\,H\,=\,\frac{f''}{(1+f'^2)^{3/2}}-\frac{1}{f\sqrt{1+f'^2}}
\end{equation}
\begin{equation}
K\,=\,-\frac{f''}{f(1+f'^2)}
\end{equation}
To determine the membrane viscous force, the tensor $K^{\alpha\beta}$ is necessary:
\begin{eqnarray}
K^{\theta\theta}\,=\,-\frac{1}{f^3\sqrt{1\,+\,f'^2}}\\
K^{\theta z}\,=\,K^{z\theta}\,=\,0\\
K^{zz}\,=\,\frac{f''}{(1\,+\,f'^2)^{5/2}}
\end{eqnarray}

%%%%%%%%%%%%%%%%%%%%%%%%%%%%%%%%%%%%%%%%%%%%%%%%%
\section*{Appendix C - basic state}
The previous quantities are calculated in the resting state, a cylinder of radius $f(z,t)=R$:
\begin{equation}
\pmb{t}^{(0)}_\theta\,=\,R\pmb{e_\theta}\: ; \:\pmb{t}^{(0)}_z\,=\,\pmb{e_z}\: ;\: \pmb{n}^{(0)}\,=\,\pmb{e_r}
\end{equation}
\begin{equation}
H^{(0)}\,=\,-\frac{1}{2R}\: ; \:K^{(0)}\,=\,0\: ; \: g^{(0)}\,=\,R^2
\end{equation}
\begin{equation}
g_{\theta\theta}^{(0)}\,=\,R^2 \;;\; g_{\theta z}^{(0)}\,=\,g_{z\theta}^{(0)}\,=\,0 \;;\; g_{zz}^{(0)}\,=\,1
\end{equation}
\begin{equation}
g^{\theta\theta,(0)}\,=\,\frac{1}{R^2} \; ;\; g^{\theta z,(0)}\,=\,g^{z\theta,(0)}\,=\,0\; ; \; g^{zz,(0)}\,=\,1
\end{equation}
\begin{equation}
K_{\theta\theta}^{(0)}\,=\,-R   \; ; \; K_{\theta z}^{(0)}\,=\,K_{z\theta}^{(0)}\,=\,0\; ;\; K_{zz}^{(0)}\,=\,0
\end{equation}
\begin{equation}
K_{\theta}^{\theta,(0)}\,=\,-\frac{1}{R}\; ;\; K_{\theta}^{z,(0)}\,=\,K_{z}^{\theta,(0)}\,=\,0\; ;\; K_{z}^{z,(0)}\,=\,0
\end{equation}
\begin{equation}
K^{\theta\theta,(0)}=-\frac{1}{R^3}; K^{\theta z,(0)}=K^{z\theta,(0)}=0;\; K^{zz,(0)}=0
\end{equation}

%%%%%%%%%%%%%%%%%%%%%%%%%%%%%%%%%%%%%%%%
\section*{Appendix D - perturbations}
Only the geometrical quantities which are necessary to the calculation of the linearized state are provided. The shape perturbation is a modulated cylinder of radius $f(z,t)=R+\delta R(z,t)$ with the condition $\mid\mid\delta R\mid\mid << R$:
\begin{equation}
\pmb{t_\theta}\,=\,\pmb{t_\theta}^{(0)}+\delta R\,\pmb{e_\theta}
\end{equation}
\begin{equation}
\pmb{t_z}\,=\,\pmb{t_z}^{(0)}+\delta R'\,\pmb{e_r}
\end{equation}
\begin{equation}
\pmb{n}\,=\,\pmb{n}^{(0)}-\delta R'\,\pmb{e_z}
\end{equation}
\begin{equation}
\delta H\,=\,H-H^{(0)}\,=\,\frac{1}{2}(\delta R''+\frac{\delta R}{R^2})
\end{equation}
\begin{equation}
\label{deltaK}
\delta K\,= K-K^{(0)}\,=K=\,-\frac{\delta R''}{R}
\end{equation}
\begin{equation}
\Delta_s \delta H\,=\,\Delta_s H\,=\,\frac{\partial^2 \delta H}{\partial z^2}\,=\, \frac{1}{2}(\delta R''''+\frac{\delta R''}{R^2})
\end{equation}
\begin{equation}
\pmb{\nabla}_s \delta\gamma\,=\,\pmb{\nabla}_s \gamma\,=\, \Big(\frac{\partial \delta\gamma}{\partial z}\Big)\,\pmb{e_z}
\end{equation}


%\balance

%\bibliography{biblio} %your .bib file
%\bibliographystyle{rsc} %the RSC's .bst file
%\footnotesize{}

\bibliography{biblio}
\end{document}